\documentclass[twocolumn,showpacs,prl,aps,superscriptaddress]{revtex4}
\usepackage{latexsym}
\usepackage{amssymb}
\usepackage[dvips]{graphicx}
\usepackage{color}

\begin{document}

\bibliographystyle{prsty}

\title{NMR evidence for a strong modulation of the Bose-Einstein Condensate in BaCuSi$_2$O$_6$}

\author{S. Kr\"{a}mer}
\affiliation{Grenoble High Magnetic Field Laboratory (GHMFL) -
CNRS, BP 166, 38042 Grenoble Cedex 09, France}

\author{R. Stern}
\affiliation{National Institute of Chemical Physics and
Biophysics, 12618,Tallinn, Estonia}

\author{M. Horvati\'c}
\author{C. Berthier}
\affiliation{Grenoble High Magnetic Field Laboratory (GHMFL) -
CNRS, BP 166, 38042 Grenoble Cedex 09, France}

\author{T. Kimura}
\affiliation{Los Alamos National Laboratory, Los Alamos NM 87545, USA}

\author{I. R. Fisher}
\affiliation{Geballe Laboratory for Advanced Materials and Department of Applied Physics,
Stanford University, Stanford CA 94305, USA}

\date{\today}

\begin{abstract}
We present a $^{63,65}$Cu and $^{29}$Si NMR study of the quasi-2D coupled spin
1/2 dimer compound BaCuSi$_2$O$_6$ in the magnetic field range 13-26 T and at
temperatures as low as 50~mK. NMR data in the gapped phase reveal that below
90~K different intra-dimer exchange couplings and different gaps
($\Delta_{\rm{B}}/\Delta_{\rm{A}}$ = 1.16) exist in every second plane along the
$c$-axis, in addition to a planar incommensurate (IC) modulation. $^{29}$Si
spectra in the field induced magnetic ordered phase reveal that close to the
quantum critical point at $H_{\rm{c1}}$ = 23.35 T the average boson density
$\overline{n}$ of the Bose-Einstein condensate is strongly modulated along the
$c$-axis with a density ratio for every second plane
$\overline{n}_{\rm{A}}/\overline{n}_{\rm{B}} \simeq 5$. An IC modulation of the
local density is also present in each plane. This adds new constraints for the
understanding of the 2D value $\phi$ = 1 of the critical exponent describing the
phase boundary.
\end{abstract}


\maketitle


The interest in Bose-Einstein condensation (BEC) has been
considerably renewed since it was shown to occur in cold atomic
gases \cite{Anderson}. In condensed matter, a formal analog of the
BEC can also be obtained in antiferromagnetic (AF) quantum spin
systems \cite{Affleck91,Giamarchi99,Nikuni00,Tanaka01} under an
applied magnetic field. Many of these systems have a collective
singlet ground state, separated by an energy gap $\Delta$ from a
band of triplet excitations. Applying a magnetic field ($H$)
lowers the energy of the $M_z= -1$ sub-band and leads to a quantum
phase transition between a gapped non magnetic phase and a field
induced magnetic ordered (FIMO) phase at the critical field
$H_{\rm{c1}}$ corresponding to
$\Delta_{\rm{min}}$-$g\mu_{\rm{B}}H_{\rm{c1}}= 0$, where
$\Delta_{\rm{min}}$ is the minimum gap value corresponding to some
\textbf{q} vector \textbf{q}$_{\rm{min}}$
\cite{Affleck91,Giamarchi99,Nikuni00,Tanaka01}. This phase
transition can be described as a BEC of hard core bosons for which
the field plays the role of the chemical potential, provided the
U(1) symmetry is conserved. Quite often, however, anisotropic
interactions can change the universality class of the transition
and open a gap \cite{Sirker04,Clemancey06,Miyahara06}. From that
point of view, BaCuSi$_2$O$_6$ ~\cite{Jaime04} seems at the moment
the most promising candidate for the observation of a true BEC
quantum critical point (QCP) \cite{Sebastian06a}. In addition,
this system exhibits an unusual dimensionality reduction at the
QCP, which was attributed to frustration between adjacent planes
in the nominally
 body-centered tetragonal structure \cite{Sebastian06}. The material also exhibits a weak orthorhombic
distortion at $\simeq$90~K which is accompanied by an in-plane IC
lattice modulation \cite{Samulon06}. This structural phase
transition affects the triplon dispersion, and the possibility of
a modulation of the amplitude of the BEC along the c-axis has been
speculated based on low field inelastic neutron data
\cite{Ruegg06}.

In order to get a microscopic insight of this system, we performed $^{29}$Si
and $^{63,65}$Cu NMR in BaCuSi$_2$O$_6$ single crystals. Our data in the gapped
phase reveal that the structural phase transition which occurs around 90 K not
only introduces an IC distortion within the planes, but also leads to the
existence of two types of planes alternating along the $c$-axis. From one plane
to the other, the intra-dimer exchange coupling and the energy gap for the
triplet states differs by 16 \%. Exploring the vicinity of the QCP in the
temperature ($T$) range 50-720~mK, we confirm the linear dependence of
$T_{\rm{BEC}}$ with $H-H_{\rm{c1}}$ as expected for a 2D BEC. Our main finding
is that the average boson density $\overline{n}$ in the BEC is strongly
modulated along the $c$-axis in a ratio of the order of 1:5 for every second
plane, whereas its local value $n$(\textbf{R}) is IC modulated within each
plane.

NMR measurements have been obtained on $\sim$10~mg single crystals of
BaCuSi$_2$O$_6$ using a home-made spectrometer and applying an external
magnetic field $H$ along the $c$ axis. The gapped phase was studied using a
superconducting magnet in the field range 13-15~T and the temperature range
3-100 K. The investigation of the FIMO phase was conducted in a 20~MW resistive
magnet at the GHMFL in the field range 22-25~T and the temperature range
50-720~mK. Except for a few field sweeps in the gapped phase, the spectra were
obtained at fixed fields by sweeping the frequency in regular steps and summing
the Fourier transforms of the recorded echoes.

\begin{figure}[t] 
\includegraphics[width=0.85\linewidth]{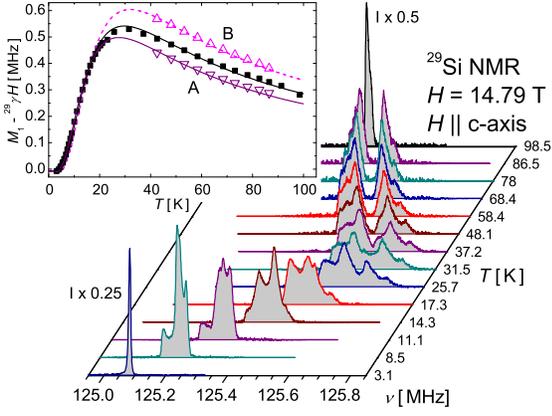}
\caption{(Color online) Evolution of the normalized $^{29}$Si NMR spectra as
a function of $T$ in the gapped phase. Below 90~K the line splits into two
components, each of them corresponding to an IC pattern. Inset: $T$ dependence
of the 1st moment (i.e., the average position) for i) the total spectra
(squares) and ii) the individual components before they overlap (up and down
triangles). The solid and dashed lines are fits for non-interacting dimers.
\label{Si_versus_T}}
\end{figure}

Before discussing the microscopic nature of the QCP, let us first consider the
NMR data in the gapped phase. The system consists of $S$ = 1/2 Cu spin dimers
parallel to the $c$ axis and arranged (at room temperature) on a square lattice
in the $ab$ plane. Each Cu dimer is surrounded by four Si atoms, lying
approximately in the equatorial plane. For Cu nuclei, the interaction with the
electronic spins is dominated by the on-site hyperfine interaction. For
$^{29}$Si nuclei both the transferred hyperfine interaction through oxygen atoms
with a single dimer and the direct dipolar interaction are important. According
to the room temperature structure $I4_{1}/acd$ \cite{Sparta04}, there should be
only one single Cu and two nearly equivalent Si sites for NMR when $H
\| c$. As far as $^{29}$Si is concerned, one actually observes a single line
above 90 K, as can be seen in Fig.~\ref{Si_versus_T}. However, below 90~K, the
line splits into two components, each of them corresponding to an IC pattern,
that is an infinite number of inequivalent sites. This corresponds to the IC
structural phase transition discovered by X-ray measurements \cite{Samulon06}.
At 3~K, when $T$ is much smaller than the gap, the spin polarization is zero
and one observes again a single unshifted line, at the frequency $\nu = \nu_0
=$~$^{29}$$\gamma H$ defined by the Si gyromagnetic ratio
$^{29}\gamma$.

\begin{figure} 
\includegraphics[width=0.85\linewidth]{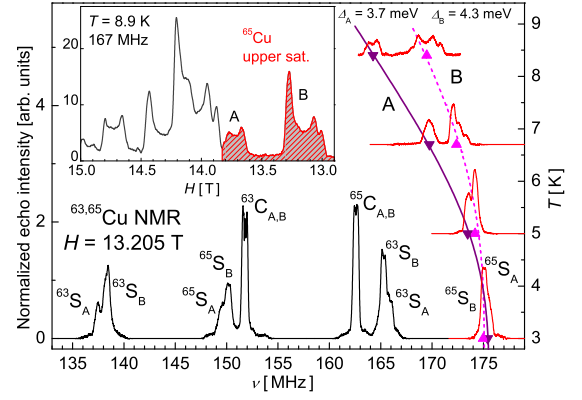}
\caption{(Color online) $^{63,65}$Cu NMR spectra of
BaCuSi$_2$O$_6$ in the gapped phase, well below the critical field. The $T$
dependence of the high-frequency ``satellite'' line clearly reveals two
different copper sites. From their shifts, the two corresponding gap values have
been determined. Inset: field sweep spectrum that reveals the IC nature of the
line shape for each of the two sites. Shading separates the contribution of the
$^{65}$Cu high-frequency satellite from the rest of the spectrum. The analysis
of the latter part confirms that the observed line shape has a pure magnetic
origin.
\label{Cu_spectra}}
\end{figure}

On the $^{63,65}$Cu NMR spectra recorded at 3~K and 13.2~T
(Fig.~\ref{Cu_spectra}), however, one can distinguish \textit{two} different Cu
sites, denoted A and B. That is, each of the 6 lines of Cu spectrum (for 2
copper isotopes $\times$ 3 transitions of a spin 3/2 nucleus) is split into
two, which is particularly obvious on the lowest frequency ``satellite''
$^{63}$Cu line. The whole spectra can be nicely fitted with the following
parameters: $^{63}\nu_{\rm{Q}}^{\rm{A} (\rm{B})}$ = 14.85 (14.14) MHz, $\eta$
= 0, and $K^{\rm{A} (\rm{B})}_{zz}$ = 1.80 (1.93) \%, where
$\nu_{\rm{Q}}$ is the quadrupolar frequency and $\eta$ the
asymmetry parameter. The $K_{zz}$ is the hyperfine shift, expected
to be purely orbital since the susceptibility has fully vanished.
On increasing $T$ the highest frequency $^{65}$Cu ``satellite''
lines of sites A and B become well separated and both exhibit a
line shape typical of an IC modulation of the nuclear
spin-Hamiltonian. Although the apparent intensities of lines A and
B look different, they correspond to the same number of nuclei
after corrections due to different spin-spin relaxation rate
1/$T_{2}$. Since the satellite NMR lines at 3~K (the lowest
temperature) are narrow, the modulation of $\nu_{\rm{Q}}$ is
negligible, meaning that the IC lineshapes visible at higher
temperature are purely magnetic. This is confirmed by the analysis
of the spectrum shown in the inset of Fig.~\ref{Cu_spectra}, which
shows that at 8.9~K the broadening of the ``central'' line is the
same as that observed on the ``satellites''.  Such a broadening
results from a distribution of local hyperfine fields: $\delta
h_{z}(\textbf{R}) = A_{zz}(\textbf{R})m_{z}(\textbf{R})$ in which
$A$(\textbf{R}) is the hyperfine coupling tensor and
$m_{z}(\textbf{R})$ the longitudinal magnetization at site
\textbf{R}. Since $\nu_{\rm{Q}}$(\textbf{R}) is not modulated by
the distortion, one expects that the modulation of $A$(\textbf{R})
is negligible too, $A$(\textbf{R})~=$A$. This means that the NMR
lineshape directly reflects the IC modulation of $m_{z}$ in the
plane.

Keeping constant the $\nu_{\rm{Q}}$ parameters obtained at 3~K,
one can analyze the $T$ dependence of the shift $K_{zz}^{\alpha
}(T)$ of each component $\alpha =$ A or B according to the formula
\begin{equation}
K_{zz}^{\alpha }(T)-K_{zz}^{\alpha }(0) =A_{zz}^{\alpha
}m_{z}^{\text{d}}(\Delta _{\alpha },H,T)/H ,
\end{equation}%
where $m_{z}^{\text{d}}$ is the magnetization of a non-interacting dimer,
$m_{z}^{\rm{d}}=g_{c}\mu _{\rm{B}}/(e^{(\Delta _{\alpha }-g_{c}\mu
_{\rm{B}}H)/k_{\rm{B}}T}+1)$ in the given $T$ range, $g_{c}=$ 2.3
\cite{Zvyagin06}, and $K_{zz}^{\alpha }$ is determined from the average line
position, i.e., the first moment. The best fit was obtained for
$\Delta_{\rm{A(B)}}$ = 3.7 (4.3) meV and $A_{cc}^{\rm{A(B)}}$ = -16.4 T/$\mu
_{\rm{B}}$. We assumed that $A^{\rm{A}}_{cc}$ = $A^{\rm{B}}_{cc}$, but the
values of $\Delta$ depend only weakly with this quantity. The values are
slightly higher than those determined by neutron inelastic scattering for
\textbf{Q}$_{\rm{min}}$ = [$\pi, \pi$] \cite{Ruegg06}, which is normal considering
our approximate description. However, the ratio
$\Delta_{\rm{B}}$/$\Delta_{\rm{A}}$
= 1.16 is in excellent agreement with the neutron result 1.15. Considering the
fact that there is no disorder in the system (as Cu lines at low $T$ are
narrow), and that X-rays did not detect any commensurate peak corresponding to a
doubling of the unit cell in the $ab$ plane, our NMR data can only be explained
if there are \textit{two types of planes} with different gap values. Looking
back at the $^{29}$Si spectra in Fig.~\ref{Si_versus_T}, one also observes just
below 90~K two well separated components, both of them exhibiting an IC pattern.
They indeed correspond to the two types of planes, as the $T$ dependence of
their positions can be well fit using values close to $\Delta_{\rm{A}}$ and
$\Delta_{\rm{B}}$ determined from Cu NMR (inset to Fig.~\ref{Si_versus_T}). This
means that the 90 K structural phase transition not only corresponds to the
onset of an IC distortion in the $ab$ plane, but also leads simultaneously to an
alternation of different planes along the $c$-axis \cite{structurefactor}, with
intra-dimer exchange in the ratio $J_{\rm{B}}/J_{\rm{A}} \cong
\Delta_{\rm{B}}/\Delta_{\rm{A}} = 1.16$

\begin{figure}[t] 
\includegraphics[width=0.85\linewidth]{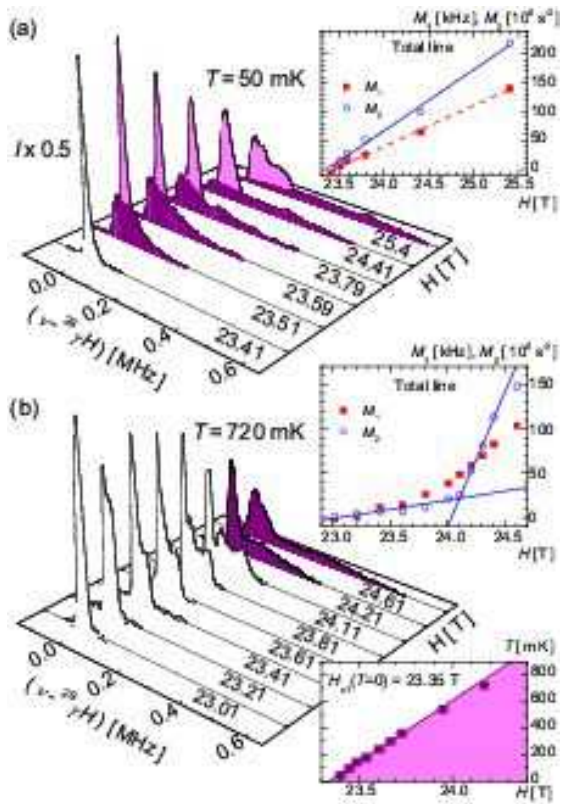}
\caption{(Color online) Evolution of the normalized $^{29}$Si
spectrum as a function of $H$ at fixed $T$. The colored spectra
correspond to the BEC. a) $T$ = 50~mK: Instead of a simple
splitting of the line as expected for a standard BEC, a complex
pattern appears, typical of an IC distribution of the local
hyperfine field. Inset: $H$ dependence of the 1$^{st}$ ($M_1$,
squares) and 2$^{nd}$ ($M_2$, circles) moment of the spectra.
$M_1$ is proportional to $m_z$ and $M_2$ to the square of the
order parameter. b) $T$ = 720~mK: The non zero magnetization
outside the BEC leads to an IC pattern for fields $H\leq
H_{\rm{c1}}(T)$, where $H_{\rm{c1}}(T)$ is determined from the $H$
dependence of $M_2$, as shown in the inset. Lower inset: $T_{c}$
is linear in $H-H_{\rm{c1}}$, as expected for a 2D BEC QCP.}
\label{Si_50mK}
\end{figure}

Let us now recall what is expected from a microscopic point of view in the
vicinity of the QCP corresponding to the onset of a \emph{homogeneous} BEC for
coupled dimer systems. As soon as a finite density of bosons $n$ is present ($H
> H_{\rm{c1}}$ = $\Delta_{\rm{min}}$/$g\mu_{\rm{B}}$), a transverse staggered
magnetization $m_{\perp}$ ($\perp$ to $H$) appears. Its amplitude
and direction correspond respectively to the amplitude and phase
of the order parameter. At the same time, the longitudinal
magnetization $m_{z}$ is proportional to the number of bosons at a
given temperature and field, this latter playing the role of the
chemical potential. Due to the appearance of a static
$\textbf{m}_{\perp}$, the degeneracy between sites which were
equivalent outside the condensate will be lifted and their
corresponding NMR lines will be split into two. To be more
specific, we consider a pair of Si sites situated in the $ab$
plane on opposite sides of a Cu dimer. Outside the condensate, and
in the absence of the IC modulation, they should give a single
line for $H$~$\|$~$c$. Inside the condensate the NMR lines of this
pair of Si sites will split by $\pm ^{29}\gamma|A_{z \perp}|
m_{\perp}$ because their $A_{z \perp}$ couplings are of opposite
sign. Obviously, observing a splitting of lines requires the
existence of off-diagonal terms in the hyperfine tensor. Such
terms are always present due to the direct dipole interaction
between an electronic and the nuclear spin, which can be easily
calculated.

Instead of this expected simple line splitting, the spectra of
Fig.~\ref{Si_50mK}a reveal a quite complex modification of the
line-shape when entering the condensate. The narrow single line,
observed at 23.41~T at the frequency $\nu_0$, which corresponds to
a negligible boson density, suddenly changes into a composite
line-shape including a narrow and a broad component. The
spread-out of the broad component increases very quickly with the
field. The width of the narrow component also increases, but at a
much lower rate. Both peculiar broadenings are related to the IC
modulation of the boson density $n$(\textbf{R}) due to the
structural modulation. To be more precise, a copper dimer at
position \textbf{R} has in total 4 Si atoms (denoted by $k$ =
0,1,2,3) situated around in a nearly symmetrical square
coordination. The absolute values of the corresponding hyperfine
couplings will thus be nearly identical, and we will also neglect
their dependence on \textbf{R}. These 4 Si sites will give rise to
four NMR lines at the frequencies $^{29}\nu_{k}(\textbf{R}) =
\nu_{0} + \nu_{1}(\textbf{R}) + \nu_{2,k}(\textbf{R})$, where
$\nu_{1}(\textbf{R}) =$~$^{29}\gamma$$A_{zz}
g\mu_{\rm{B}}n$(\textbf{R}) and $\nu_{2,k}(\textbf{R})
=$~$^{29}\gamma$$A_{z\perp} m_{\perp}(\textbf{R})
\cos(\phi-k\pi/2)$. Note that $\nu_{2,k}$ only exists when the
bosons are condensed, that is when there is a transverse
magnetization $m_{\perp}$ pointing in the direction $\phi$. In a
uniform condensate $m_{\perp}$ is proportional to $\sqrt{n}$ near
the QCP, since the mean field behavior is valid in both, 2D and
3D. We assume that only the amplitude of the order parameter is
spatially modulated, and that $m_{\perp}(\textbf{R}) \propto
\sqrt{n(\textbf{R})}$. The line shape is the histogram of the
distribution of $^{29}\nu_{k}(\textbf{R})$, convoluted by some
broadening due to nuclei -- nuclei interaction.

Three quantities can be derived from the analysis of NMR lines at
fixed $T$ values and variable $H$: the average boson density
$\overline{n}(H, T)$, the field $H_{\rm{c1}}(T)$ corresponding to
to the BEC phase boundary, and the field dependence of the BEC
order parameter (for $T$ close to zero). The average number of
bosons $\overline{n}$ per dimer is directly proportional to the
first moment $M_{1}$ (i.e., the average position) of the line:
$M_{1} = \int_{-\infty}^{\infty} (\nu-\nu_{0}) f(\nu){\rm d}\nu
=$~$^{29}\gamma$$A_{zz} g\mu_{\rm{B}}\overline{n}(H, T)$, where
the line shape $f(\nu)$ is supposed to be normalized. The second
moment (i.e., the square of the width) of the line $M_{2} =
\int_{-\infty}^{\infty} (\nu-\nu_{0}-M_{1})^{2} f(\nu){\rm d}\nu$
has two origins: the broadening due to the IC distribution of
($n$(\textbf{R})-$\overline{n}$), and that due to the onset of
$m_{\perp} \propto \sqrt{n(\textbf{R})}$ in the condensate. When
increasing $H$ at $T \simeq 0$, the condensation occurs as soon as
bosons populate the dimer plane. This is observed in the inset of
Fig.~\ref{Si_50mK}a at $T$~=~50~mK. Both $M_{1}$ ($\overline{n}$)
and $M_{2}$ ($m_{\perp}$) vary linearly with the field and the
extra{\-}polation of $M_{2}$ to zero allows the determination of
$H_{\rm{c1}}$ at 50~mK. For higher temperatures a thermal
population of bosons $\overline{n}$ exists and increases with $H$
before entering the BEC phase. As a result both $M_{1}$ and
$M_{2}$ increase non-linearly with $H$, as shown in the upper
inset of Fig.~\ref{Si_50mK}b. However, the increase of $M_{2}(H)$
shows two clearly separated regimes and allows the determination
of $H_{\rm{c1}}(T)$ as the point where the rate of change of
$M_{2}(H)$ strongly increases due to the appearance of
$m_{\perp}$. Applying this criterion to all temperatures, we were
able to determine the field dependence of $T_{\rm{BEC}}$ (lower
inset of Fig.~\ref{Si_50mK}b) and define precisely the QCP at
$H_{\rm c1}$ = 23.35~T. In agreement with the torque measurements
\cite{Sebastian06}, we find a linear field dependence. This is the
signature of a 2D BEC QCP, where $T_{\rm{c}} \propto
(H-H_{\rm{c1}})^{\phi}$ with $\phi= 2/d$ and $d$~=~2
\cite{Batista}.

\begin{figure}[t] 
\includegraphics[width=0.85\linewidth]{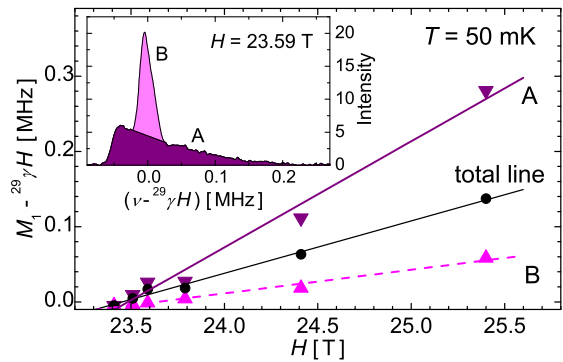}
\caption{(Color online) Using a simple decomposition of the
spectra into two components as shown in the inset, we determined the
1$^{\rm{st}}$ moments of the $^{29}$Si lines corresponding to the different
types of planes A and B. From the slopes of their field dependence, the ratio of
the average boson density is found equal to
$\overline{n}_{\rm{A}}/\overline{n}_{\rm{B}} \simeq 5$.}
\label{BEC_split}
\end{figure}

This analysis, however, does not take into account the specificity
of the line shapes, which are related to the existence of two
types of planes with different energy gaps. A careful examination
of the spectra clearly reveals that they correspond to the
superposition of two lines exhibiting different field dependence
at fixed $T$ value. For sake of simplicity, we have made a
decomposition only for the spectra at 50~mK, as shown in the inset
to Fig.~\ref{BEC_split}. Clearly, one of the components remains
relatively narrow without any splitting, whereas the other
immediately heavily broadens in some sort of triangular line
shape. The field dependence of $M_{1}$ of the two components,
shown in Fig.~\ref{BEC_split}, reveals that they differ by a
factor of 5. This is attributed to the difference by a factor of 5
in the corresponding average populations of bosons. If there were
no hopping of bosons between A and B planes, the B planes should
be empty for the range of field such that $\Delta_{\rm{A}} <
g\mu_{\rm{B}}H < \Delta_{\rm{B}}$. Although the observed density
of boson is finite in the B planes, it is strongly reduced, giving
rise to a strong commensurate modulation of $\overline{n}$ along
the c-axis. According to \cite{Sebastian06}, the hopping along the
$c$-axis of bosons in the condensate is forbidden by the
frustration, and can only occur as a correlated jump of a pair.
However, this argument does not take into account the IC
modulation of the boson density.

In conclusion, this NMR study of the 2D weakly coupled dimers
BaCuSi$_2$O$_6$ reveals that the microscopic nature of the BEC in
this system is much more complicated than first expected. Two
types of planes are clearly evidenced, with different intra-dimer
$J$ couplings and a gap ratio of 1.16. Close to the QCP we
observed that the density of bosons, which is IC modulated within
each plane, is reduced in every second plane along the $c$-axis by
a factor of $\simeq$ 5. This provides new constraints for the
understanding of the quasi-2D character of the BEC close to the
QCP.

\begin{acknowledgments}

We thank S.E. Sebastian, C.D. Batista and T. Giamarchi for discussions. Part of
this work has been supported by the European Commission through the EuroMagNET
network (contract RII3-CT-2004-506239), the Transnational Access
- Specific Support Action (contract RITA-CT-2003-505474), the Estonian
Science Foundation (grant 6852) and the NSF (grant DMR-0134613).

\end{acknowledgments}



\begin{thebibliography}{10}
\bibitem{structurefactor}
This does not introduce any superstructure peak along (0,0,$l$) in X-ray
experiments, since the unit cell already contains four planes along the
$c$-axis. Only the form factor, which has not been studied in details below
90~K, should be slightly affected.
\end{thebibliography}
\end{document}